\documentstyle[11pt,fleqn,leqno,epsfig,wrapfig]{article}
\begin{document}

\title{An extension of ``Popper's experiment'' can
test interpretations of quantum mechanics}
\author{        
        R.Plaga
        \\Max-Planck-Institut f\"ur Physik (Werner-Heisenberg-Institut)
        \\ F\"ohringer Ring 6
        \\D-80805 M\"unchen, Germany
        \\email:plaga@mppmu.mpg.de
        }
\maketitle
\begin{abstract}
\noindent
Karl Popper proposed a way to test
whether a proposed relation of a quantum-mechanical state 
to perceived reality in
the Copenhagen interpretation (CI) of quantum mechanics
- namely that the
state of a particle is merely an expression of ``what is known''
about the system - is in agreement with all experimental facts.
A conceptual flaw in Popper's proposal
is identified and
an improved version of his experiment 
(called {\it ``Extension step 1''}) - which fully serves
its original purpose - is suggested.
The main purpose of this paper is to suggest to
perform this experiment.
The results of this experiment predicted under the alternative 
assumptions that the CI - together with the above 
connection of the state function with reality -
or the ``many-worlds'' interpretation 
(MWI) is correct are shown to be identical.
Only after a further modification 
(called{\it ``Extension step 2''}) -
the use of an ion isolated from the macroscopic environment 
as particle detector -
the predictions using the respective interpretations become
qualitatively different.
This is because 
``what is known'' by a human observer H
can fail as a basis for the prediction
of the statistical distribution of measurement 
results within the MWI in special cases:
The temporal evolution of a
system un-entangled with H - like the isolated ion - can 
depend on another system's state components
that are entangled with states ortogonal to H. Thus - within
the CI - for H they are ``known not to exist''.
Yet H can infer their existence by studying the evolution
of the ion. 
\end{abstract}
\noindent
\section{Introduction - aims and plan of the paper}
\label{sec_intro}
A considerable number of ``contemporary'' (i.e. writing
within the last decade) authors \footnote{
They will be represented below by B.-G.Englert-M.O.Scully-
H.Walther, L.Mandel, A.Peres, R.Peierls, W.G.Unruh and A.Zeilinger}
share a ``standard'' understanding of the foundations of
quantum mechanics that has its
roots in the ``Copenhagen interpretation'' formulated by 
Bohr, Heisenberg, Pauli and others.
A central idea of this approach 
is that the quantum-mechanical
state function\footnote{Below
the denotations ``state function'' is used interchangeably
with ``wavefunction'' and ``density matrix''.} 
faithfully and {\it exclusively}
represents our {\bf knowledge} of a system.
I will try to formulate the standard
interpretation (called ``CI'' for brevity below)
more precisely in section \ref{cisum} and clarify its relation to
the historical Copenhagen interpretation in section \ref{cop_his}.
\\
``Popper's thought experiment'',
was proposed by the philosopher K.R.Popper
\cite{popper,popperbook,combo}. 
It aims to {\it experimentally test} the relation
of a quantum-mechanical state to observed reality
as described in the previous paragraph. 
Popper's work is reviewed in section \ref{sec_ori}.
His own expectation for the outcome of the 
experiment and his ideas for the interpretation of quantum
mechanics are not the subject of the present paper.
\\
Popper's proposal
contains two flaws, one technical (section \ref{flaw1}) the other
conceptual (section \ref{flaw2}). Recently Kim and Shih
found an ingenious method to avoid the former
(section \ref{kimshih}). 
The latter can be avoided 
by slightly modifying Kim and Shih's experimental set-up,
as proposed here for the
first time ({\it Extension step 1}, sections \ref{critical},
\ref{outcome}).
Thus modified - but not in its original form -  
Popper's experiment will serve its original purpose.
It is the primary aim of this paper to propose
to perform this technically feasable (see section
\ref{tech_diff}) experiment.
\\
Section \ref{mwipop} discusses the predictions of the ``many-worlds
interpretation'' (MWI) 
- an alternative to the CI (section \ref{mwi}) -
for this set-up and concludes that they are the same as the ones
of the CI.
Based on similar results
some authors concluded that
all alternative interpretations 
{\it which leave the standard 
mathematical quantum-mechanical formalism unaltered} 
(like CI and MWI) give rise to
the same phenomenological predictions under all
circumstances\cite{zeilingerfp,tegmark}\footnote{No 
general proof for this claim is
given in these references.}.
If this were true the question which of these alternative
interpretations is ``correct'' would be a purely philosophical one.
It is a secondary aim of this paper to prove this contention
wrong.  
\\
To this end I discuss a further extension of
``Popper's experiment'' 
in section \ref{sec_popmod},
and describe the expected results of this
experiment both in the CI and MWI ({\it Extension step 2}).
The respective results are qualitatively different.
Section \ref{tech_diff} discusses the considerable technical difficulties 
when experimentally realising this modified experiment.
Finally, in section \ref{summary} I summarise the main conclusions
of the present paper.
Sections in small font provide mainly quotes
to substantiate my understanding of the ``standard view'', they
can be skipped without damage to the understanding of the paper.
\section{The ``Copenhagen'' interpretation
as formulated by modern authors}
\label{sec_ci}
In this section I summarise 
the ``standard'' interpretation of quantum 
mechanics given by many modern authors in order to understand
its predictions for Popper's experiment.
This ``standard view'' 
is based on the historical ``Copenhagen interpretation'' 
(and therefore abbreviated as ``CI'' in this article) but
the authors quoted below (with the exception of Zeilinger) 
do not to explicitely call it like this. More than in 
deep insights of a philosophical nature (which was a central
point for the founding fathers) they seem to be interested
primarily in a simple and unique set of rules that make the mathematical
formalism of quantum mechanics work in accordance with experience.

{\footnotesize Typical statements are:
\\
``This minimalistic interpretation of state vectors
is ``{\it forced upon us} by the abundance of empirical
facts that show that quantum mechanics works'' (Englert et al.
in Ref.\cite{englert},p.328,
sec. VI, l.14). ``Minimalistic'' 
is meant in the sense of a renunciation
of an understanding of the state function as something existing
independently of our minds.
\\
``...when you refer to the Copenhagen interpretation what you really
mean is quantum mechanics.'' (Peierls in Ref.\cite{peierlsgia}, p.71, l.8).}
\subsection{Summary of CI interpretation}
\label{cisum}
If at a time t=0
a ``measurement'' is performed on a complete set of commuting observables
A,B,C... of a physical system,
the results obtained (i.e. non-degenerate 
eigenvalues a$_n$,b$_n$,c$_n$ ...)
define a unique eigenvector $\Psi$ of all observables.
$\Psi$(t=0) is called the ``state vector'' and is a complete description
of the system's state.
This operation is called ``preparation'' of the system.
The temporal evolution of $\Psi$ is defined by deterministic
equations of motion, e.g. Schr\"odinger's equation for nonrelativistic
electrons or QED for photons.
When an observable A is measured at a later time t $>$ 0 the probability
P to obtain an eigenvalue a$_n$ is 
\begin{equation}
P(a_n)=\sum_{i=1}^{g_n} |<u^i_n|\Psi>|^2
\end{equation}
here $g_n$ is the degree of degeneracy of a$_n$ and the $<$u$^i_n|$
are the eigenvectors associated with the eigenvalue a$_n$.
The state function is thus the
unequivocal basis for prediction of statistical
results of future measurements.
\subsubsection{Relation of the state function to observed
reality}
\label{statedef}
``Measurement'' is defined here as 
``knowledge obtained by observations''
(Peierls in Ref.\cite{peierls}, p.778, l.34). 
``The most fundamental statement of
quantum mechanics is that the state vector ... represents our {\it
knowledge} of the system''\cite{peierls}(Peierls's emphasis, 
p.778, l.20).
This definition of the state vector is emphasised by all 
authors. {\footnotesize Englert et al. \cite{englert} (p.328,IV.,l.7)
explain their ``minimalistic interpretation''
of the state vector -
mentioned above - like this:
``The state vector $\Psi$(x) serves the sole purpose
of summarising concisely our knowledge about the ... system;
in conjunction with the known dynamics it enables us to make 
correct predictions about the statistical properties of future
measurements.''
Unruh\cite{wald}(p.883,left column,l.22) describes the state vector 
in its ``usual epistemological role'' as 
``a device within the theory to incorporate our knowledge of the
world, without it in itself corresponding to anything in the real 
world.'' Zeilinger\cite{zeilinger} writes
(p.3,l.10): ``In the Copenhagen interpretation the
state function is only our way of representing that part of our knowledge
of the system which is needed for calculating future probabilities 
for specific measurement results.''}
Mandel\cite{mandel} (p.S279,right column,l.13) - based on
the results of quantum-optical experiments -
and Peres\cite{perespriv} define the 
nature of the ``knowledge'' more precisely:
Mandel states:
``the state reflects not what is actually known about
the system, but what is knowable, in principle, with the help
of auxiliary measurements that do not disturb the original
experiment.'' In order to not contradict the previous authors and
to avoid counterfactual reasoning I suggest to clarify 
Mandel's wording to:
``...not {\it only} what is actually known about
the system, but {\it also} what is knowable...''.
Peres makes clear that the knowledge is not the knowledge of
a system existing independently of the observer when he replaces
``knowledge of the system'' by ``knowledge of the preparation of the
system''.
There are eminent contemporary advocates
of the ``standard interpretation'' (e.g. Asher Peres in Ref.\cite{peres})
who avoid to use the above understanding of ``state function''
all together, but
they do not criticise it as incorrect.
\subsubsection{Whose knowledge is relevant?}
\label{whose}
Because information cannot spread with velocities exceeding c
``what is knowable, in principle with the help
of auxiliary measurements'' about a certain system
can be different for observers at space-like distances.
Thus the state function of a given system can be different
for these observers.
Which state function then has to be used for the prediction of the statistical
properties of future measurements?
Peierls\cite{peierls} gave the following answer:
``Each (observer) has to use his or her state function.''(p.779,
right column,l.11). In other words:
{\it The knowledge relevant for experimental outcomes
is the one of the observer 
performing a measurement of the state function representing
the knowledge.} This answer 
seems very natural, because any alternative
potentially violates causality.
\subsection{Is this ``modern Copenhagen interpretation'' the same
as the one of the founding fathers?}
\label{cop_his}
{\footnotesize 
A complete analysis of the Copenhagen interpretation's history
is beyond the scope of the present paper which tries
to make possible an 
experimental test of the CI as defined by contemporary physicists.
However it is important to ask
whether the principal architects
of the Copenhagen interpretation (Bohr, Heisenberg and Pauli among
others) shared the unequivocal view expressed above
by representative modern authors that the quantum-mechanical
state function is merely an expression of our knowledge of the
system. D.Mermin - who has studied the original writings in
great depth (especially those of Bohr) -
concluded recently\cite{mermin} (p.3,l.4):
``...those who maintain, unambiguously with Heisenberg and
presumably with Bohr, that the state function is nothing more than
a concise encapsulation of our knowledge.''
Pauli repeatedly 
stressed the observer dependent character
of physical laws that follows from such a view of the state function;
e.g. in a letter to Bohr he wrote\cite{zeilinger}(in ref.28):
``...I see the unpredictable change of the state through the
individual observation ... as a rejection of the idea of the detachment
of the observer of the course of physical events outside himself.''
I conclude that the writings of the founding fathers at least do not
contradict the modern CI view.}     
\section{The ``many-worlds'' interpretation}
\label{mwi}
The conceptually simplest alternative interpretation
to the CI  
is Everett's ``many-worlds'' interpretation\cite{everett,
zeh,zurek,deutsch,tegmark}
that proposes to plainly accept the mathematical formalism as
an exact description of an objective observer-independent reality.
In particular the state function exists as an element of
reality completely independent of any observer.
It can be shown that a necessary consequence of these
strong but natural assumptions
is that after each quantum measurement ``branches'' with
independently existing observers form - each of which observes ones
of the quantum mechanically possible results.
Proponents of the MWI suggest to take these ``bizarre''
consequences from the equations of quantum mechanics serious, in the
same way as Einstein's ``bizarre'' conclusions from
Maxwell's equations (such as time dilation) were taken serious
before their direct experimental confirmation.
Understanding whether CI or MWI establishes the correct
correspondence between perceived world and theory 
is one of the most profound questions of physics.
As an example within the MWI quantum mechanics is a deterministic
theory whereas within the CI chance plays a fundamental role
in nature. 
\section{Popper's experiment}
\label{sec_pop}
\begin{figure}[hbt]
\centering

\epsfig{
file=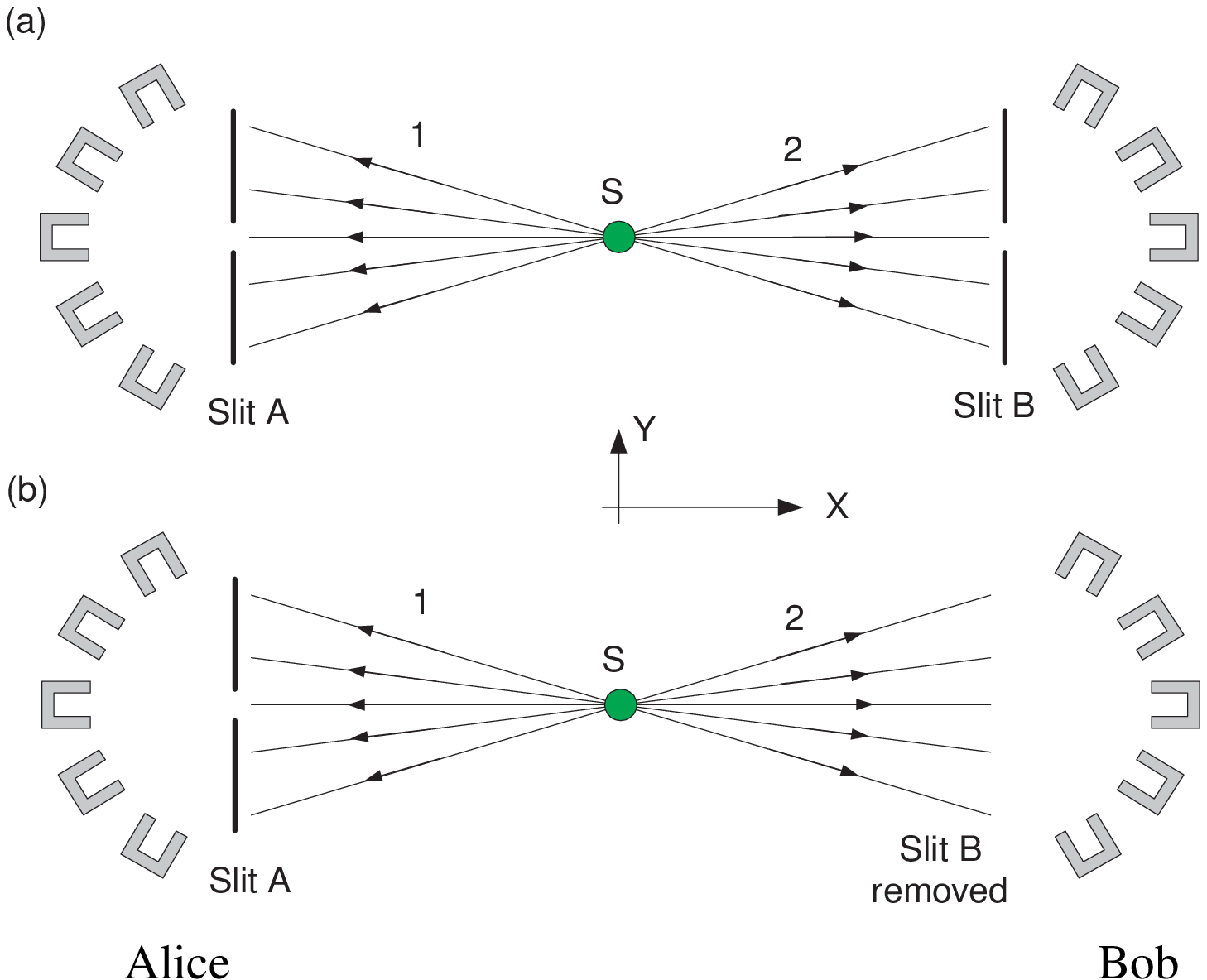,width=12cm,height=7cm,clip=,angle=0}
\caption{ \label{popperidea} \it 
Popper's experiment as originally proposed (from Ref.\cite
{kim}). A pair of entangled particles
is emitted from a point source S. In panel A slits are placed on both
sides of S and diffraction takes place as a result of the localisation
at the respective slits. All counters in the figure fire in repetitions
of the experiment.
Panel B is the experiment proposed: Particle
1 is localised in a slit. As a results of momentum conservation
particle 2 is also localised in spite of the absence of a slit
at its side\cite{peresaj}. Does the knowledge of its position, gained by 
Alices's
observation of particle 1 lead to an increased momentum scatter,
i.e. ``virtual diffraction''? Do all counter on the right side continue
to fire in repetitions of the experiment?}
\label{fig1b}
\end{figure}

\subsection{The original proposal (fig. \ref{fig1b})}
\label{sec_ori}
Let a 
point source S - located at rest in the middle on the connecting 
line between two slits A and B -
decay into two entangled particles, e.g. positronium
decaying into two photons, or a photon being down converted into
an ``idler'' and a ``signal'' photon. The position in y 
of the particle propagating towards slit A (particle 1)
is determined
with the slit A. If one of the detectors behind slit A
fires, the position in y is known with a precision
$\Delta$y = d$_{\rm slit}$, where d$_{\rm slit}$ is the slit width.
According to the uncertainty relation the momentum of this
particle can then only be known with a precision
$\Delta$ p$_{\rm y}$ = $\hbar$/d$_{\rm slit}$. In other
words: diffraction takes place at slit A.
The dimensions
of the setup are chosen such, that - as a consequence of the
increased momentum spread - all counters on the left
side (rather than only the one located behind 
slit A on the conecting line with slit B)
fire with equal probability when the particle passes through
the slit.
\\
In order to guarantee momentum conservation the particles
are emitted back to back\cite{peresaj}. As a consequence when
particle 1 passed through slit A, particle 2 must have
passed through slit B. Both particles experience an
increased momentum scatter (diffraction)
and all counters on both sides
fire in repetitions of the experiment (panel a. in fig \ref{fig1b}).
\\
Popper's question was: what happens if one removes slit B 
(panel b. in fig \ref{fig1b})?
The expectation of the CI {\bf seems to be} (here 
- as an introduction to the basic question - I faithfully repeat Popper's
argument which is {\bf not} entirely correct, 
see sections \ref{flaw1} and
\ref{flaw2} below):
As soon as the experimenter ``Alice'' 
- reading the detectors behind slit A -
knows that particle 1 passed through slit
A (i.e. after it passed the x position of slit A) she knows
that particle 2 is localised within slit B; slit B still
acts as a ``virtual slit''.
Because the state function is ``what is known about the 
preparation of the system''
one has to use the localised state of particle 2 to predict
the statistics of the
results of the measurement with counters on the right side.
The ``preparation'' is Alice's test whether particle 1 passes
slit A.
Thus an increased momentum spread is expected at the virtual
slit for all those particles 2 {\it which are detected
in the detectors at the right in coincidence with
a detection of particle 1 after slit A.} This
coincidence requirement is assumed hold in all arguments of this
manuscript.
``Virtual diffraction'' is expected to take place
in exactly the same way as before with the 
presence of a physical slit B, and
all detector behind slit B can fire.
Mere knowledge can change the unitary evolution of photon 2.
It is the {\it change of unitary evolution caused by state
preparation}, which makes
Popper´s experiment qualitatively different from previous similar
experiments.
\\
Popper thought that this result  
would {\bf not} obtain in a really performed experiment,
for reasons that are not of interest here.
\subsection{First technical flaw in the original proposal - the 
position determination 
via entanglement is less precise than the direct one}
\label{flaw1}
\begin{figure}[hbt]
\centering
\begin{tabular}{c@{\qquad}c}
\mbox{
\epsfig{file=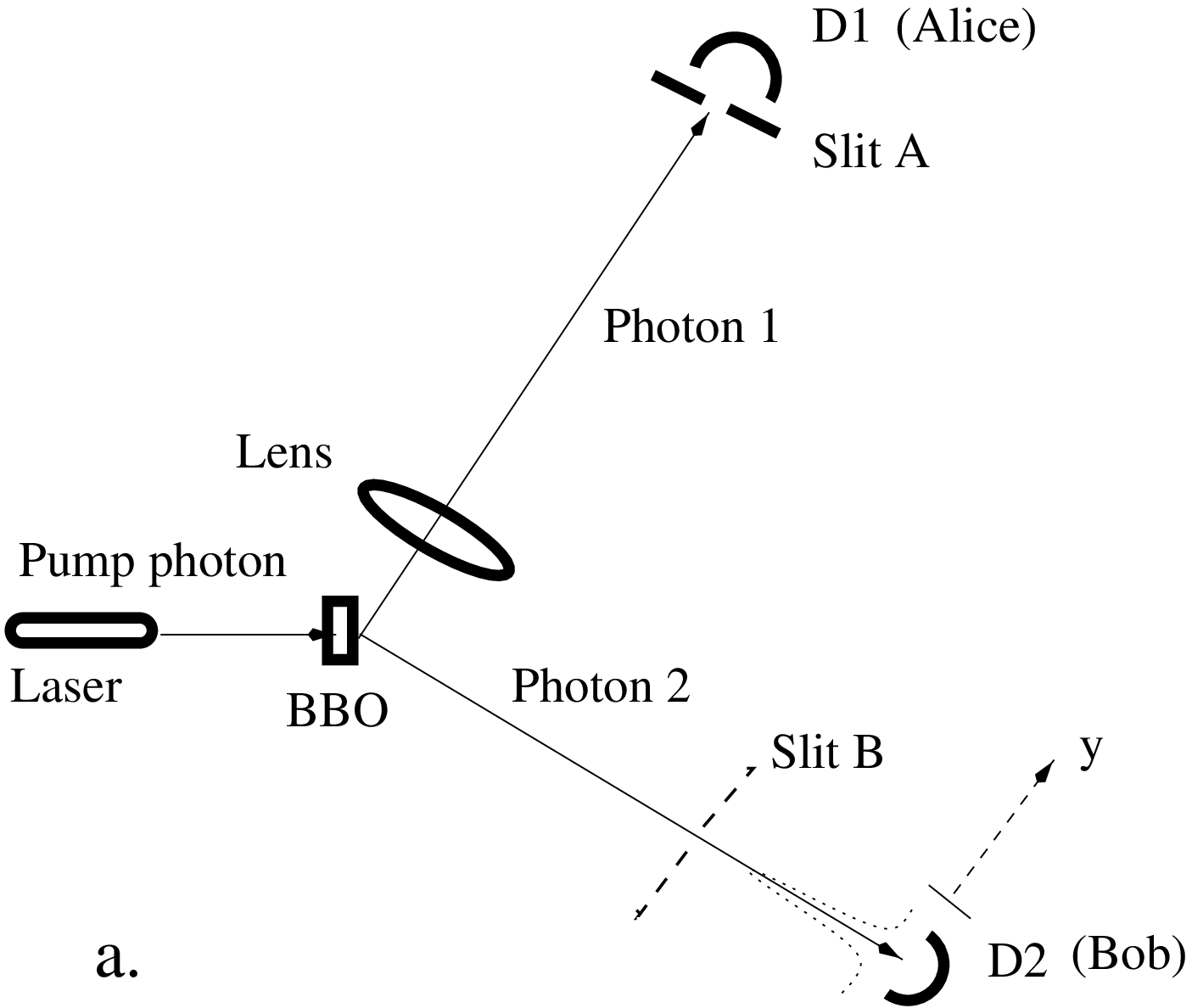,width=7cm,height=7cm,clip=,angle=0}}&
\mbox{\epsfig{file=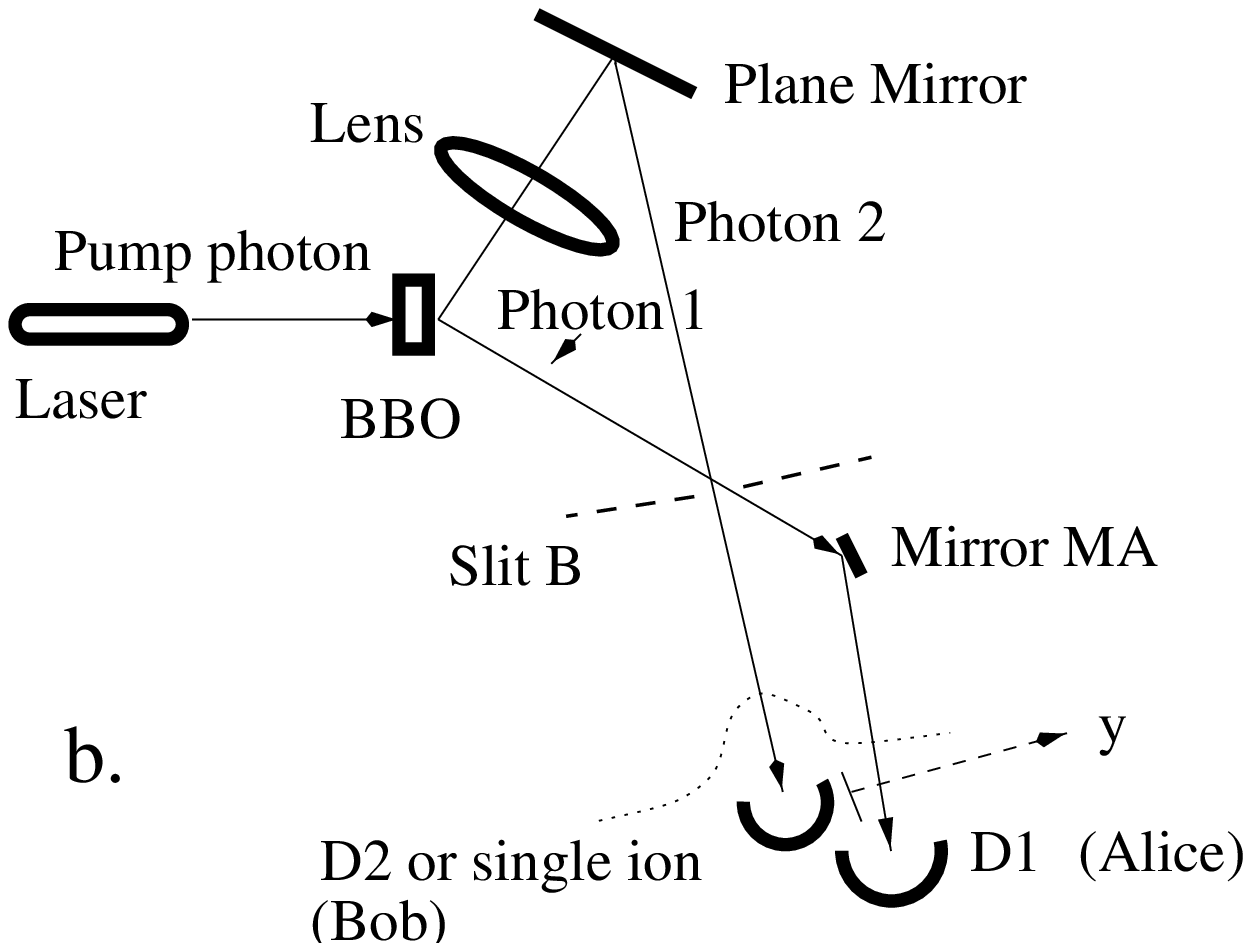,width=7cm,
height=7cm,clip=,angle=0}}
\end{tabular}
\caption{ \label{results} \it 
a. the experiment of Kim and Shih\cite{kim}; b. Extension step 1 
- version of their experiment in which an increased
momentum spread is expected within the CI as a result of
the virtual localisation of photon 2 (figures adapted 
in simplified form from
figure 2 of their paper). The ``virtual slits'' are shown with
dashed lines. The dotted lines in front of the detector D2
symbolise the distribution of photon 2 in y-direction
with a small (a.) and large (b.) momentum spread. If a single ion
- which was isolated before D1 clicked - 
is used instead of a macroscopic detector for D2 (``Extension step 2'')
the MWI predicts
a small momentum spread also in case b.}
\label{fig1}
\end{figure}

In 1987 Collett and Loudon\cite{collett,poppernat,collett2} 
pointed out an elementary
oversight in Popper's setup of fig \ref{fig1b}.
The source S has to obey the uncertainty relation
and therefore if one
localises S with a small spatial region 
(as necessary for a point source)
its y-momentum component becomes uncertain. The particles
are then no longer always emitted back to back. They showed that
as a result of this, for any source size
a localisation of particle 1 within slit A
does not imply a localisation of particle 2 within the
``virtual slit B'', but only within a much broader region.
The authors did not imply that this fact makes Popper's
test impossible in principle, some years later Collett published
a proposal for a completely
different experiment which would realise Popper's
idea\cite{storey}.
\subsubsection{The solution of Kim and Shih (fig. \ref{fig1}a)}
\label{kimshih}
Kim and Shih (KS)\cite{kim} recently demonstrated
the feasibility of experimentally performing ``Popper's
experiment'' for the first time.
In their experiment (fig. \ref{fig1}a, a simplified
version of fig.2 in their publication) two entangled 
photons ``1'' and ``2'' 
(this they call a ``biphoton'') are produced via parametric
down conversion of a pump 
photon in the beta barium borate ``BBO'' crystal.
Photons 1 and 2 are entangled because they obey ``phase-matching 
conditions'', i.e. their energies and momenta have to add
up to the pump photon's values of these parameters.
\\
Via the introduction of an optical lens in the path of photon 1, KS
ensured that if
photon 1 is known to be localised
within slit A (i.e. detector D1 clicked),
the phase matching conditions
localise the quantum state of
photon 2 within slit B of equal size.
This is true even if it is not known from where in BBO crystal
the photons were emitted.
Even when choosing the emission region in the BBO crystal sufficiently
large to avoid Collett and Loudon's criticism,
the localisation of particle 2 within a virtual slit B -
after it is known that particle 1 passed through slit A - can
be ensured with arbitrary precision.
\\
Detector D2 is located at a distance $a$
behind a virtual slit B. It is movable in y direction - perpendicular to
the direction of photon propagation x - and 
measures the y-momentum spread p$_y$ of the quantum state of
photon 2 with total momentum $p$ via the relation p$_y$=y$\times p$/$a$.
A quantum state of photon 2
localised within a ``virtual slit'' of size d$_{\rm slit}$
has a momentum 
spread of $\Delta$p$_{\rm slit}$ $\approx$ $\hbar$/d$_{\rm slit}$
according to Heisenberg's uncertainty relation.
Experimentally KS did {\bf not} find this, i.e. the momentum
spread of photon 2 was not increased according to the uncertainty
relation.
\subsection{Second conceptual flaw in the original proposal -
the localisation via entanglement is not knowable to the observer
performing the momentum-spread determination}
\label{flaw2}
When trying to find the correct prediction of the CI
for both Popper's and KS's setup we run into the 
problem explained in section \ref{whose}. 
When the observer ``Bob'' behind the
virtual slit B performs the measurement of particle 2's momentum spread
it is in principle still not knowable for him
- for obvious causality reasons -
whether particle 1 passed though slit A.
According to the definition in section \ref{statedef}
his state function is thus still the original, unlocalised one.
According to the discussion in section \ref{whose}
- in spite of the fact that the observer ``Alice'' near slit A
already knows that the particle passed the slit -
{\it one expects no increased momentum spread in the CI}.
{\bf This} is the reason why the experimental result of KS is accordance
with the CI and why one expects {\bf no} increased momentum spread
also for Popper's original setup.
This is in agreement with the point of view of
Peres\cite{perespopper} and KS who argued that no increased
momentum spread is expected for Popper's original setup.
The uncertainty relation for a particle 
has to hold for the state function of the observer measuring
its position and momentum i.e. the one relevant for the prediction
of experimental results.
Therefore particle 2, which
remains unlocalised for Bob, shows no increased momentum
scatter without violating the uncertainty relation, in accordance
with the conclusion of KS.

\subsubsection{Fixing the last flaw - 
the ``Extension step 1'' form of Popper's experiment (fig.\ref{fig1}b)}
\label{critical}
We search for an experiment in which
the following two conditions are fulfilled:
\\
1. The click of detector 1 has to signal the
localisation of particle 2 within the virtual slit B.
\\
2. At time when particle 2 reaches detector D2\footnote{
Because of the small relative velocities of Bob,Alice and
the experimental setup the ususal nonrelativistic notion of temporal
coincidence can be used.}
Bob has to be at a space-time point
on the past light-cone of detector 1's click.
\\
If conditions 1 and 2 are both fulfilled the click 
of detector 1 makes it knowable to Bob
- the observer performing the measurement and whose knowledge
is therefore relevant to define the state function
(see section \ref{whose}) - 
that the particle is localised within the virtual slit B. 
According to the quotes in section \ref{statedef} {\it by definition}
the state function - which Bob must use to predict the
statistical properties of future measurements -
then {\it is} spatially localised in this slit.
The standard time evolution of quantum mechanics 
now predicts an enhanced momentum spread of this particle
(i.e. the occurence of virtual diffraction).
{\it Only} if both conditions at the beginning of this section 
are fulfilled the CI is tested in the way envisaged by Popper.
\\
Formally the initial photon state
$\Psi_{\rm photon}$  (the ``biphoton'' state of the 
entangled photon pair) is given as the superposition
of two states with labels a and b corresponding to
the possible measurement results\cite{kim}:
\begin{equation}
|\Psi_{\rm photon}>(\rm{initial}) = |\Psi_{\rm a}> +  |\Psi_{\rm b}> =
|\Psi_{\rm i1}>|\Psi_{\rm i2}> + |\Psi_{\rm o1}>|\Psi_{\rm o2}>
\label{entangled}
\end{equation}
Here label 1 (2) correspond to photon 1 (2)
and i (o) to a state in which the photon 
localised inside (outside)
the physical (for particle 1) or  virtual (for particle 2) slit position.
To find the particle state after the measurement of particle 1
(the ``click'' of detector 1)
I project the particle state onto the eigenstate
of the position:
 \begin{equation}
|\Psi_{\rm photon}>(\rm{after\, click \, of \, detector \, 1}) = 
|\Psi_{\rm i1}>|\Psi_{\rm i2}> 
\label{projected}
\end{equation}
Such state clearly evolves into the well known diffraction pattern 
of a single slit with size d
in the far field (where the localisation probability is
proportional to sin$^2$(x)/x$^2$
with x=$\phi$ k $\cdot$ d, where k is the wave number of the particle
and $\phi$ the angle with respect to the slit center).
\\
Fig. \ref{fig1}b shows a setup slightly modified from the one of KS
to fulfill both conditions (``Extension step 1'').
The basic setup, the focal length of the lens and the optical
path-lengths ``BBO to slit A and B''
remains exactly the same as in KS's performed
experiment.
The physical slit A and the virtual slit B are exchanged
together with detectors and particle labels.
A plane mirror is introduced behind the lens which deflects
the optical path of photon 2.
Instead of slit A a plane mirror MA which size is equal
to the slit width of slit A. Only photons reflected
from this mirror can hit D1, therefore a click in D1
signals that the photon was localised within the slit width.
\\ 
Condition 1 is fulfilled because the optical path
``BBO - lens - plane mirror - virtual slit B'' has the
same length as the one ``BBO - Mirror MA - D1'', i.e. a
click of D1 signals the localisation of photon 2 within
the virtual slit B.
To fulfill condition 1 quantitatively 
it is important to ensure that photons {\bf not} hitting
mirror MA are not absorbed before photons reflected are absorbed
by D1. Otherwise a close inspection of the absorbing matter
would make it ``in principle knowable'' that the photon 1
hit MA {\bf before} photon 2 reaches slit B.
\\
Condition 2 is fulfilled because  
the optical path ``BBO - lens - plane mirror -
D2'' is longer than the one ``BBO - Mirror MA - D1 - D2''.
This means that the result of the position measurement
of photon 1 can be known by Bob - the observer measuring the
momentum spread of photon.
\subsection{Possible outcomes of Popper's experiment in the
``Extension step 1'' form}
\label{outcome}
It is of great importance to actually perform Popper's experiment
with ``Extension step 1''.
Storey et al.\cite{storey} - in a different realisation
of Popper's experiment - expect that virtual diffraction will be
observed. The MWI interpretation predicts the same result
as discussed in the next subsection.
Should an experimental realisation of ``Extension step 1''
show that {\it no} virtual diffraction occurs the relation between
``quantum-mechanical state'' and ``
observed reality'' proposed in section 2.1.1 and
2.1.2 would be put into doubt. The proposed ``Extension step 2''
of section 5 would then not be of interest. 
The further discussion of an extended experiment below
{\it assumes} that virtual diffraction will be found in ``Extension step 1''.

\subsection{Analysis of Popper's experiment within the MWI}
\label{mwipop}
The initial state $|\Psi_{\rm initial}>$ after down conversion
of all experiments discussed here can be written as:
\begin{equation}
|\Psi_{\rm initial}> = |\Psi_{\rm photon}> \otimes |M> \otimes |A>
\label{initial}
\end{equation}
Here $|$M$>$ stands for the state of the 
``rest of the universe'' 
excluding $|\Psi_{\rm photon}>$ and $|A>$
but including the measurement device and
possibly a human observer. $|$A$>$ is an ion isolated from
the environment in a particle trap. $|$A$>$ is not used in
KS's and Popper's 
experiment ``Extension step 1'' 
but will be of importance in the ``Extension step 2''
version discussed below in section \ref{sec_popmod}.
When the position of
photon 1 is determined,
the biphoton interacts with $|$M$>$ and entanglement occurs.
The entanglement then spreads with a velocity presumably
not very different from c in the environment. 
The final state after entanglement is given as\cite{knight}:
\begin{equation}
|\Psi_{\rm final}> = |\Psi_{\rm lab}> \otimes |A> =
(|\Psi_{\rm i1}>|\Psi_{\rm i2}> 
|M_{\rm a}> + |\Psi_{\rm o1}>|\Psi_{\rm o2}> |M_{\rm b}>)
\otimes |A>
\label{final}
\end{equation}
The MWI assumes that a quantum state
is an objective description of reality.
Further the localisation of the photon happens upon entanglement
of $|$M$>$ with $|\Psi_{\rm photon}>$
(an objective physical process) rather than when its position
becomes ``knowable to an observer'' as in the CI.
It thus accepts expression (\ref{final})
as plain reality,
i.e. after entanglement (which ``for all practical
purposes'' is irreversible due to the complexity of $|$M$>$)
there are - among other things - two 
human observers in different ``branches'' described
by $|M_{\rm a}>$ and $|M_{\rm b}>$.
These observers do not ``see'' each other
because $|M_{\rm a}>$ and $|M_{\rm b}>$
do not interfere, due to their entanglement
with the states $|\Psi_{\rm i1}>|\Psi_{\rm i2}>$ and
$|\Psi_{\rm o1 }>|\Psi_{\rm o2}>$, respectively, that
are orthogonal in Hilbert space.
This is completely analogous to amplitudes of a
single atom that do not interfere with each other
in a ``which-way'' experiment, due to their 
respective entanglement with orthogonal states\cite{scully}.
After the photon and  $|$M$>$ are entangled, the observer
described by  $|M_{\rm a}>$ only interacts
with $|\Psi_{\rm i2}>$. She then measures an increased momentum spread
according to the uncertainty relation.
\\
A determination of the momentum spread of photon 2
at a space-time point before entanglement 
of $|$M$>$ with $|\Psi_{\rm photon}>$ took place
corresponds to KS's experiment; performed
after this entanglement took place it corresponds to the
Popper's experiment ``Extension step 1''.
\\
In the framework of the MWI
the fact that KS found a momentum spread of photon 2
corresponding to an unlocalised state allows to draw
the following nontrivial conclusion ``CONCL''.
{\it Let the two components
$|\Psi_{\rm i1}>$ and $|\Psi_{\rm o1}>$ 
of a particle 1 be entangled with two 
components $|\Psi_{\rm i2}>$ and $|\Psi_{\rm o2}>$
of a particle 2 that are orthogonal
to each other.
In spite of this fact (which precludes the possibility
of any interference between the components of particle 1)
a system $|$M$>$ - initially not entangled with
either particle - that measures the momentum uncertainty
of particle 1 finds the value corresponding
to the total state $|\Psi_{\rm i1}>$ + $|\Psi_{\rm o1}>$.}

\section{Popper's experiment ``Extension step 2'':
an isolated system as detector 
(fig.\ref{fig1}b with a ``single
ion'' as detector D2)}
\label{sec_popmod}
The described absence of 
interference and the linearity of quantum mechanics
- which precludes any other influence of 
the ``branches'' $|M_{\rm a}>$
and $|M_{\rm b}>$ on each other - has 
led to the view that MWI and CI give
rise to the same phenomenology in principle
\cite{tegmark}. 
In this case the decision between CI and MWI 
would forever remain a matter of philosophical taste.
This view is erroneous: this equivalence {\bf only} 
holds in the approximation
that the measurement device is always entangled with
its environment. To demonstrate this 
I now discuss a further modification
of Popper's experiment (``Extension step 2'').
Let us assume in the following that photon 1 was found
within the slit (first term in eq.(\ref{final}) $|$M$_{\rm a}>$ describes
the subjective observer).
Consider the ion in the state $|$A$>$ in expression (\ref{initial})
and (\ref{final}).
Its isolation from the environment
(e.g. in a particle trap in a good vacuum)
prevented its entanglement with the environment
for the duration of the experiment.
Now this particle is employed as the photon detector D2 for
photon 2, i.e. if the ion is found excited the detector has ``fired''.
\\
Within the MWI $|$M$>$ in eq.(\ref{initial}) and
$|$A$>$ in eq.(\ref{final}) have exactly the same status.
In particular conclusion ``CONCL'' above holds. We only
replace $|\Psi_{\rm i2}>$  and $|\Psi_{\rm o2}>$ by 
$|\Psi_{\rm i2}>$ $|M_{\rm a}>$ and $|\Psi_{\rm o2}>$ $|M_{\rm b}>$,
respectively,
two components which continue to be orthogonal.   
Therefore $|$A$>$ finds the same as $|$M$>$ before, i.e.
{\it measuring with the ``single-ion detector'' one 
finds the small momentum spread as in KS's experiment
even after $|$M$>$ became entangled with the biphoton}.
If this result is ascertained with the ion detector at a y-position
where it does not interact with the photon (but would, if the momentum
spread were increased to $\Delta$p$_{\rm slit}$) $|$A$>$ does {\it not} 
get entangled in the course of the measurement process via the photon. 
\\
I summarise the expectation for Popper's experiment
in the ``Extension step 2'' form for CI and MWI:
\\
CI:
\\
Bob (the observer after slit B) measures an increased
momentum spread as expected from the localisation of particle 2
in the virtual slit, which is known to him.
\\
This prediction also holds for all interpretations 
or modifications of quantum mechanics in
which a physical collapse of particle 2's wavefunction 
explains its localisation.
\\
MWI:
\\
If, and only if, Bob uses a detector which
is un-entangled with the biphoton
to measure the momentum uncertainty of particle 2 
he will find a localised particle 2 with no increased momentum spread.
\subsection{Technical possibilities to perform the 
proposed experiments}
\label{tech_diff}
KS's experiment with the ``Extension step 1'' could
be readily performed with relatively small changes in their setup.
``Extension step 2'' is more difficult, but not out of reach
of present technology.
The degree of isolation necessary to prevent an entanglement
of a single particle with a macroscopic environment
is nothing special for single particles;
it has to be ensured in all experiments
in which interference phenomena 
of atoms or neutrons are to be experimentally studied.
The observable excitation of single ions - necessary
to employ a single ion as photon detector - is
routinely practised by some quantum-optics groups
(e.g. \cite{walther}). Of course such a detector has
an exceedingly low efficiency.
However, the necessity to produce
a very large number of ``biphotons'' for such
a purpose (the single-ion detector
has a very low efficiency) does not preclude
to perform the experiment with state of the art technology.
\section{Summary}
\label{summary}
After fixing two important shortcomings (i.e. in the setup
of Kim and Shih with {\it Extension step 1}) Popper's experiment
becomes a test of a non-trivial 
tenet of the Copenhagen interpretation (CI):
that the state function represents our knowledge of a system. 
The main proposal of this paper is to perform this experiment.
An experimental result confirming the above tenet
would be strikingly counterintuitive. 
The very different many-worlds interpretation (MWI)
predicts the same result for this experiment.
Assuming the confirming result in an 
actual experiment additional
{\it Extension step 2} could be performed 
- in which the predictions made using CI
versus MWI are qualitatively different.
Because - in my opinion - both interpretations are logically 
tenable presently, it is not possible to predict the outcome
of an experiment performed in this way.
 
{\bf Acknowledgements} I thank Y.Kim for a 
helpful correspondence about the
technical details of his experiments and M.Collett and A.Peres
for clarifications about ``Popper's experiment''. S.Pezzoni, H.D.Zeh 
and two referees contributed many helpful criticisms on the manuscript. 
\vskip -2.0 cm

\end{document}